\documentclass[aps,twocolumn,showpacs,floatfix,nobibnotes]{revtex4}
\usepackage[dvips]{graphicx}

\begin{document}
\title{Flexomagnetic effect in Mn-based antiperovskites}

\author{Pavel Lukashev}
\affiliation{Department of Physics, University of Nebraska, Omaha,
Nebraska 68182}

\author{Renat F. Sabirianov}

\altaffiliation[Also at ]{Nebraska Center for Materials and
Nanotechnology, University of Nebraska, Lincoln, Nebraska 68588}

\affiliation{Department of Physics, University of Nebraska, Omaha,
Nebraska 68182}

\date{\today}

\begin{abstract}

We report appearance of the net magnetization in Mn-based
antiperovskite compounds as a result of the external strain gradient
(\textit{flexomagnetic effect}). In particular, we describe the
mechanism of the magnetization induction in the Mn$_{3}$GaN at the
atomic level in terms of the behavior of the local magnetic moments
(LMM) of the Mn atoms. We show that the flexomagnetic effect is
linear and results from the non-uniformity of the strain, i.e. it is
absent not only in the ground state but also when the applied
external strain is uniform. We estimate the flexomagnetic
coefficient to be 1.95 $\mu_{B}$\AA \hspace{0.05cm}.
We show that at the moderate values of the strain gradient ($\sim
0.1\%$) the flexomagnetic contribution is the only non-vanishing
input to the induced magnetization.

\end{abstract}
\maketitle

\begin{figure}[t]
\begin{center}\vskip 0.5 cm
\includegraphics[width=6cm]{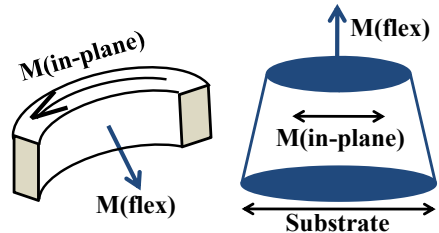}
\caption{Schematic picture of the systems under strain gradient with
possible orientations of the induced magnetization.
\label{flex-schematic}}
\end{center}
\end{figure}

Magneto-mechanical coupling in crystals has many practical
applications, such as in sensors, magnetic recording devices, etc.
The correlation between external strain and induced magnetization is
in principal significant for the systems with different
dimensionality, i.e. bulk, thin film, and nano configurations. Yet,
the correlation between the strain gradient and induced
magnetization is especially important in nanostructures and thin
film heterostructures because of the large surface-to-volume ratio
which may result in a large surface tension due to the structural
distortions caused by lattice mismatch, external stress, etc. For
example the strain gradient may play a significant role when thin
film is epitaxially grown on a substrate with slightly different
lattice parameters. On the other hand, in problems concerning bulk
structures the strain gradient is small and as a result has only
negligible contribution. This is so because of the dimensional
scaling inherent in the very definition of the strain gradient, i.e.
the decrease of the characteristic length of the system, $x$ results
in increase of the $\Delta x/x$ ratio. Fig.~\ref{flex-schematic}
schematically shows possible geometries of the systems under
external strain gradient: bent nano-wire and nano-pill grown on a
substrate. The "in-plane" and "out-of-plane" orientations of the
induced magnetization shown on Fig.~\ref{flex-schematic} will be
explained later in the text.

The magneto-mechanical coupling is phenomenologically described by
adding additional terms to the expression for the thermodynamic
potential, i.e. free energy, proportional to the product of the
magnetic field component and the conjugate terms involving
mechanical strain:

\begin{eqnarray}
\label{free-en-piezo} F_{magn} = -\lambda_{i,jk}H_{i}\sigma_{jk} -
\mu_{i,jk}H_{i}\sigma_{jk}^{2} -
\nu_{ijkl}H_{i}\frac{\partial\sigma_{jk}}{\partial x_{l}}; \nonumber \\
\hspace{0.2cm} i,j,k,l = 1,2,3 \hspace{0.5cm}
\end{eqnarray}
where $\lambda_{i,jk}$ is the piezomagnetic tensor, $H_{i}$ is the
\textit{i}-th component of the magnetic field ($i=x,y,z$),
$\sigma_{jk}$ is the elastic stress tensor, $\mu_{i,jk}$ is the
magneto-elastic tensor, $\nu_{ijkl}$ is the 4-rank tensor
(\textit{flexomagnetic tensor}), and
$\frac{\partial\sigma_{jk}}{\partial x_{l}}$ is the strain gradient.
For example, piezomagnetic (magnetostrictive) properties of certain
antiferromagnetic (AFM) materials are reflected in a term linear
both in the magnetic field and in the elastic stress tensor
\cite{Landau-Lif}. For the bulk structures the last term on the
r.h.s. of (\ref{free-en-piezo}) is usually omitted because of its
negligible contribution, yet in nanostructures and/or thin film
heterostructures it plays a significant role.

By taking partial derivative of (\ref{free-en-piezo}) w.r.t.
magnetic field component, $H_{i}$ we get the net magnetization in
the system, which may be linear w.r.t. the strain (i.e.
piezomagnetic effect), quadratic (second order magneto-elastic
effect), and proportional to the strain gradient:

\begin{eqnarray}
\label{flex-mom} M_{i} =
\underbrace{\lambda_{i,jk}\sigma_{jk}}_{piezo} +
\underbrace{\mu_{i,jk}\sigma_{jk}^{2}}_{mag-elast} +
\underbrace{\nu_{ijkl}\frac{\partial\sigma_{jk}}{\partial
x_{l}}}_{flex}; \hspace{0.2cm} \nonumber \\ i,j,k,l = 1,2,3
\hspace{0.5cm}
\end{eqnarray}

It is important to emphasize that the symmetries of the
$\lambda_{i,jk}$ and $\mu_{i,jk}$ tensors are different from the
symmetry of the $\nu_{ijkl}$ tensor in (\ref{flex-mom}), which means
that the induced magnetization will have three contributions in
principal distinguishable by symmetry arguments only. Moreover, it
is possible that because of the crystal and magnetic symmetry the
linear piezomagnetism will be absent in the system with the nonzero
strain gradient induced magnetization. We define the contribution to
the $M_{i}$ from the third term on the r.h.s. of Eq.
(\ref{flex-mom}) as \textit{flexomagnetic (FlM) effect}, i.e. strain
gradient induced magnetization.

To the best of our knowledge the term \textit{flexomagnetoelectric
effect} was first coined by Bobylev and Pikin in their study of the
correlation between elastic and electro-magnetic properties of
nematic liquid crystals \cite{Bobylev-Pikin}. As a sidenote we
emphasize here that in their work Bobylev and Pikin discussed the
"reverse" flexomagnetoelectricity, i.e. the re-orientation (which
they called \textit{flex}) of the molecules under external electric
and magnetic fields. Much work was done in the past to investigate
the correlation between electric polarization and external strain
gradient, at both experimental and theoretical level
\cite{Tagantsev}$^{,}$ \cite{Indenbom}$^{,}$ \cite{Wenhui}. Yet, the
correlation between magnetic behavior of the system and the gradient
of the strain was not the subject of mainstream research. One of the
reasons for the insufficient study on this matter is the obvious
complexity of the problem, - as opposed to the flexoelectric effect,
where the electric polarization directly correlates with the atomic
displacements, the FlM effect is indirect, i.e. it results from the
re-orientation of the atomic spins following the atomic
displacements (because of the exchange interaction). As a result,
for the FlM effect one has to consider not only the crystal but also
the magnetic structure and symmetry. Systems of interest must
satisfy certain conditions, such as they have to be non-magnetic in
ground state, and at the same time they have to exhibit strong
magneto-elastic coupling. Comparatively well known manifestation of
these properties is the piezomagnetic effect, i.e. induction of a
spontaneous magnetic moment in the system under mechanical strain.
In what follows we present the results of our study of the external
strain gradient induced magnetization mechanism.

\begin{figure}[h]
\begin{center}\vskip 0.5 cm
\includegraphics[width=5cm]{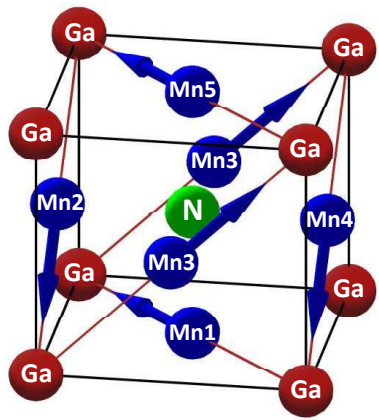}
\caption{(color online) Ground state of the antiperovskite
Mn$_{3}$GaN unit cell: non-collinear $\Gamma$$^{5g}$ magnetic
structure. Local magnetic moments are shown by blue arrown on Mn
atoms. \label{Gamma-5g}}
\end{center}
\end{figure}

To understand the mechanism of flexomagnetism at the atomic level we
perform first principles study for the Mn$_{3}$GaN under strain
gradient. The choice of the material is based on the intriguing
magneto-mechanical coupling mechanism in Mn$_{3}$GaN. In its ground
state Mn$_{3}$GaN is non-magnetic with non-collinear $\Gamma$$^{5g}$
structure (in the classification of Bertaut et al. \cite{Bertaut}),
i.e. the Mn LMMs on the (111) plane form clockwise or
counterclockwise configurations, such that the spin moments in the
plane are compensating each other. The atoms of Mn, Ga and N form an
antiperovskite (AP) crystal structure, and the lattice constant of
the primitive 5-atom cell of Mn$_{3}$GaN is 3.86 \AA.
Fig.~\ref{Gamma-5g} shows the unit cell of the Mn$_{3}$GaN in the
ground state.

In our recent work \cite{Lukashev} we have shown that the
application of the external stress to the Mn-based AP compounds
results in appearance of the non-zero magnetic moment, i.e. these
compounds exhibit \textit{piezomagnetic} properties. This can be
explained from the symmetry viewpoint, i.e. as a result of the
applied biaxial strain the symmetry of the system reduces from the
trigonal space group \textit{P\={3}1m} to the orthorhombic
P$m^\prime m^\prime m$ ferromagnetic space group. Because of this
transition some of the symmetry operations are not compatible
anymore with the new structure. The appearance of the net
magnetization in the system under external stress is due to the
rotation of the LMMs of the Mn atoms from their equilibrium
positions. After rotation the LMMs in the (111)-plane become
inequivalent and do not compensate each other anymore. The
piezomagnetic effect is linear and magnetization reversal is
potentially possible upon reversal of the sign of the strain
(compressive to tensile or vice versa). In the present work we
examine the magnetic behavior of the Mn$_{3}$GaN under strain
gradient (\textit{flexure}), i.e. we look at the third term on the
r.h.s. of Eq. (\ref{flex-mom}).

We use projector augmented wave (PAW) method by Bl\"{o}chl
\cite{Bloechl}, implementation of PAW by G. Kresse and D. Joubert
\cite{Kresse} in the Vienna \textit{ab} initio simulation package
(VASP) code within a Perdew-Burke-Ernzerhof (PBE) generalized
gradient approximation \cite{PBE} of the density functional theory
(DFT). We use a 3x12x6 k-point sampling and the Bl\"{o}chl's
tetrahedron integration method \cite{Bloechl1}. We set the
plane-wave cut-off energy to 300 eV and we choose the convergence
criteria for energy of 10$^{-5}$ eV.

\begin{figure}[h]
\begin{center}\vskip 0.5 cm
\includegraphics[width=6.5cm]{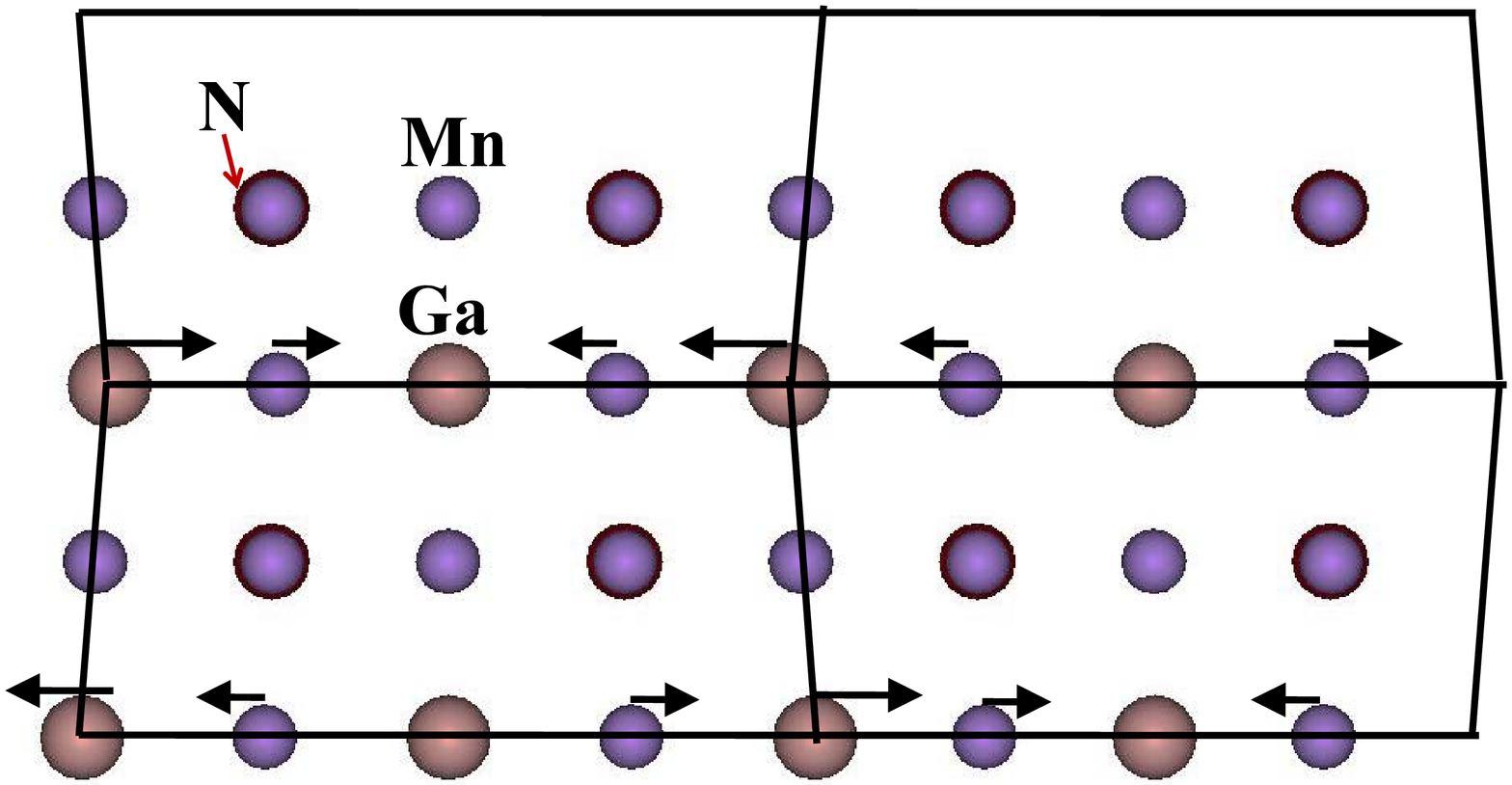}
\includegraphics[width=6cm]{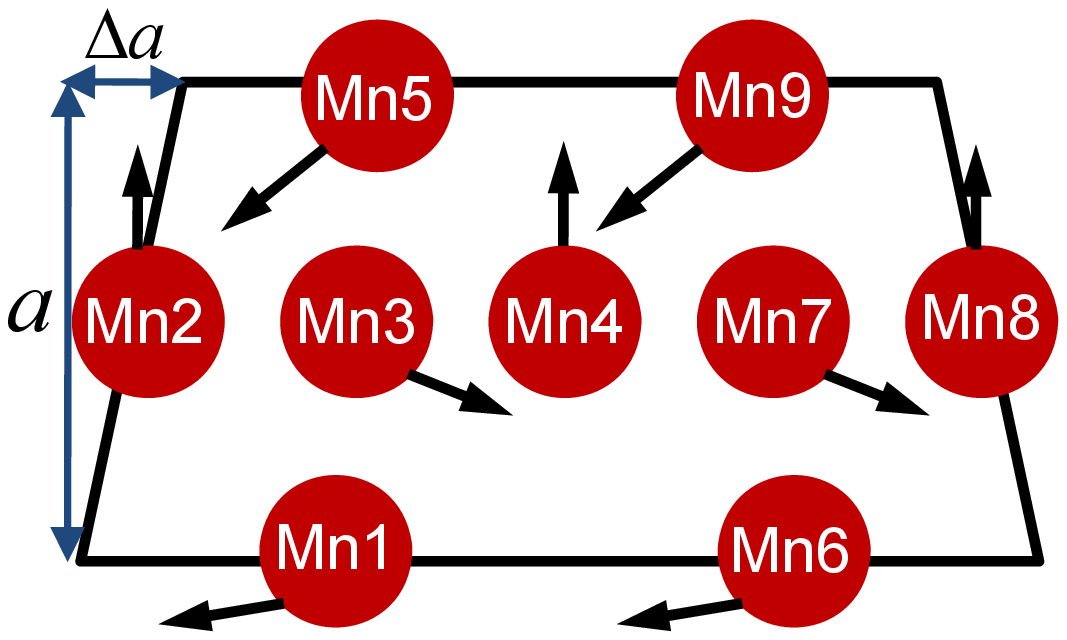}
\caption{(color online) Top panel: Mn$_{24}$Ga$_{8}$N$_{8}$ cell:
4-domain structure. Mn atoms - blue, Ga atoms - grey, N atoms - dark
red (almost invisible behind Mn atoms). The directions of the atomic
shifts are schematically shown by black arrows. Bottom panel: Single
domain in Mn$_{24}$Ga$_{8}$N$_{8}$ under strain gradient.
\label{4-domains}}
\end{center}
\end{figure}

Accurate electronic structure calculations require periodic boundary
condition. The strain gradient breaks the periodicity of the
primitive unit cell, therefore to construct a model with
translational symmetry we have to take larger cell. For Mn$_{3}$GaN
the smallest possible configuration to retain the translational
symmetry under external strain gradient consists of the 8 primitive
cells of Mn$_{3}$GaN (4x2x1 cell configuration). To simulate the
strain gradient we introduce small relative atomic shifts of Ga and
Mn atoms in a way, which splits our 40-atom cell in 4 domains (see
Fig.~\ref{4-domains}, top panel). In each of the four domains we
have strain gradient, $\Delta a/a$ (schematically shown on the
bottom panel of the Fig.~\ref{4-domains}), and 4-domain
configuration satisfies the periodic boundary condition. The initial
orientation of the LMMs forms non-collinear $\Gamma$$^{5g}$
structure. In our model we have "flat" triangular LMM configuration
(Fig.~\ref{4-domains}, bottom panel), while on Fig.~\ref{Gamma-5g}
LMMs are on the (111) plane. Both magnetic orientations are
energetically indistinguishable, and we pick the "flat" system for
simplicity. Table~\ref{tab:table1} summarizes the values for the
atomic shifts and the strain gradients for the
Mn$_{24}$Ga$_{8}$N$_{8}$ cell.

\begin{table}
\begin{tabular}{|c|c|c|c|}
  \hline
  Step & Mn (\AA) & Ga (\AA) & Flex (\%) \\ \hline
  1 & 0.01868 & 0.03736 & 0.242 \\ \hline
  2 & 0.03736 & 0.07442 & 0.484 \\ \hline
  3 & 0.05589 & 0.11179 & 0.724 \\ \hline
  4 & 0.07450 & 0.14900 & 0.965 \\ \hline
  5 & 0.09264 & 0.18528 & 1.200 \\ \hline
  6 & 0.11132 & 0.22234 & 1.442 \\ \hline
\end{tabular}
\caption{Atomic shifts of Mn and Ga atoms and strain gradients:
Mn$_{24}$Ga$_{8}$N$_{8}$.}
 \label{tab:table1}
\end{table}

\begin{figure}[h]
\begin{center}\vskip 0.5 cm
\includegraphics[width=6cm]{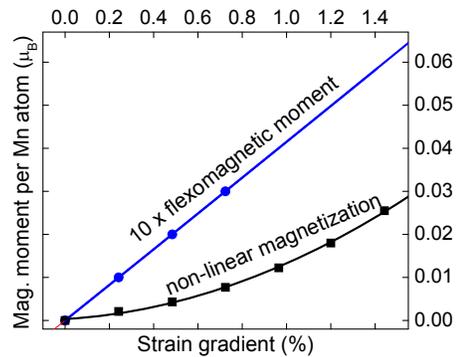}
\caption{(color online) Linear out-of-plane ($\times$10) and
non-linear in-plane induced net magnetization of Mn atom as a
function of the strain gradient. \label{magn-vs-flexure}}
\end{center}
\end{figure}

Next we relax the LMMs of the Mn atoms but we keep the atomic
positions fixed to make sure that the crystal structure does not
relax back to the unstrained ground state.
Fig.~\ref{magn-vs-flexure} shows results of our calculations for the
magnetization per Mn atom as a function of the strain gradient.
There are two different contributions: the blue line with circles
represents out-of-plane magnetization (multiplied by 10) which is a
direct result of the flexure. If the strain applied to the cell is
uniform then from the symmetry arguments it is clear that only
in-plane magnetization will appear. The black line with squares
represents in-plane non-linear contribution to the magnetization
which results from the in-plane rotations of the LMMs of the Mn
atoms (shown on the bottom panel of the Fig.~\ref{4-domains} by
black arrows). Important feature of these two mechanisms is that at
the moderate values of the strain gradient the dominant contribution
comes from the linear flexomagnetic magnetization, while the
non-linear in-plane contribution is vanishingly small. This
correlation changes at the higher values of the gradient but the
dominant nature of the linear contribution at the moderate gradient
values is important for the practical applications where the regular
values of the strain gradient are of the order of $\sim
0.1\div0.2\%$.

\begin{figure}[h]
\begin{center}\vskip 0.5 cm
\includegraphics[width=6.5cm]{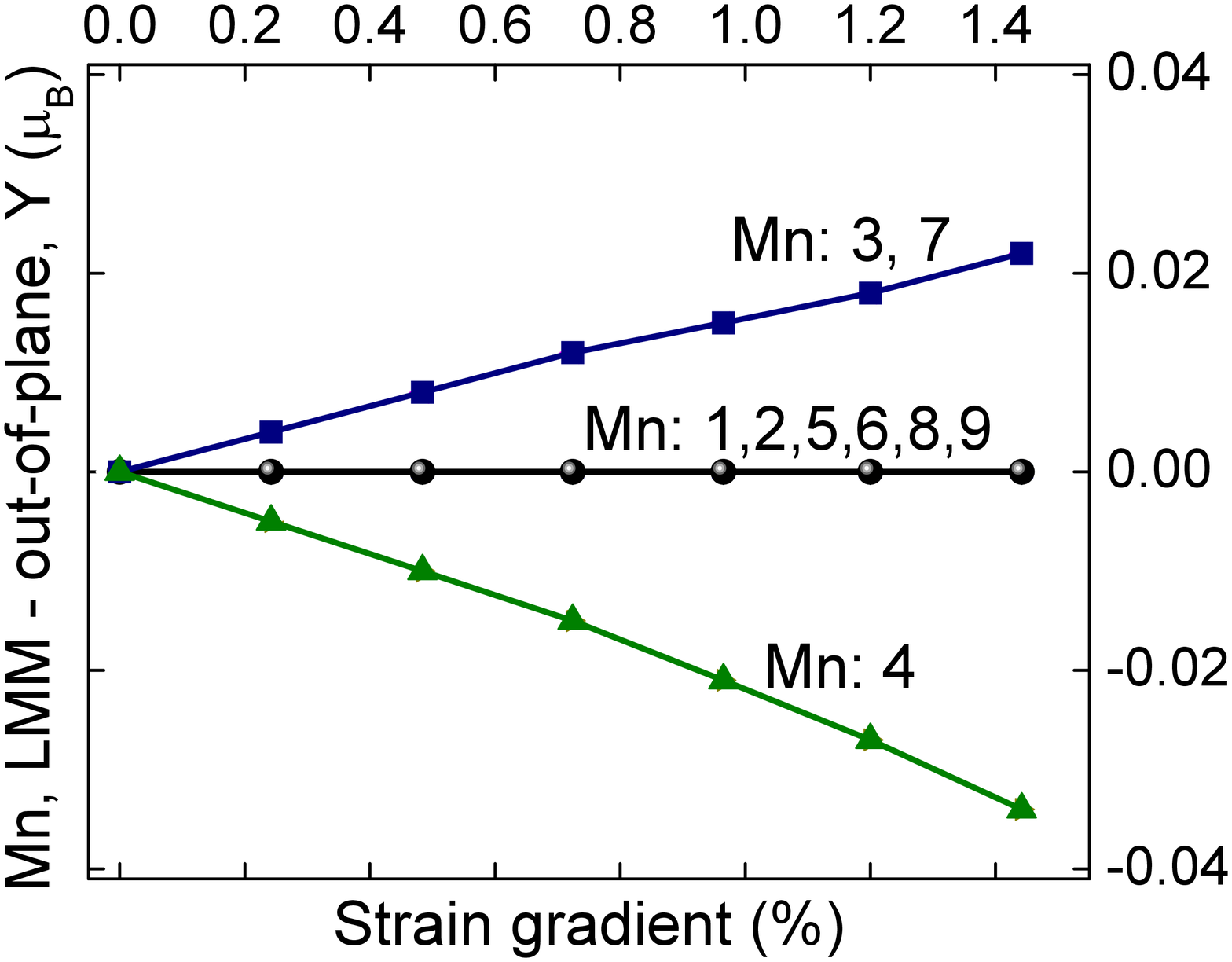}
\caption{Out-of-plane magnetic moment vs. strain gradient.
\label{out-of-plane}}
\end{center}
\end{figure}

\begin{figure}[h]
\begin{center}\vskip 0.5 cm
\includegraphics[width=6.5cm]{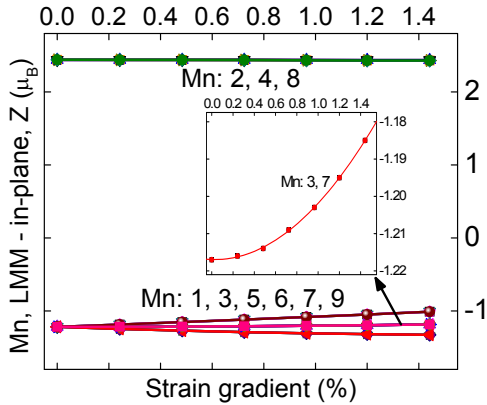}
\caption{In-plane magnetic moment vs. strain gradient.
\label{in-plane}}
\end{center}
\end{figure}

To understand better the nature of these two contributions we
examine the behavior of in-plane and out-of-plane components of the
LMMs of Mn atoms within one domain. Figures~\ref{out-of-plane} and
~\ref{in-plane} show our results. The out-of-plain contribution
comes from the Mn atoms 3, 4 and 7 (see Fig.~\ref{4-domains}, bottom
panel). At the same time the in-plane rotations of the Mn LMMs
(atoms 3 and 7) show distinct non-linear feature, which contributes
to the non-linear in-plane magnetization shown on
Fig.~\ref{magn-vs-flexure}. The appearance of the out-of-plane
component is not observed if the applied strain is uniform.
Therefore, the appearance of the linear out-of-plane magnetization
is purely strain gradient related effect and its mechanism is
different from the one responsible for the in-plane magnetization
induction. Below we present our phenomenological arguments on the
mechanism of the FlM effect at the atomic level and estimate the
value of the FlM coefficient.

Qualitatively, the out-of-plane rotations of the LMMs can be
explained by the frustration of the spin orientations. Because of
the inequivalent strain on different lattice sites the ground state
nearest neighbor distances between Mn atoms change in non-uniform
way. This is schematically shown on the bottom panel of the
Fig.~\ref{4-domains} where for example the Mn1$\leftrightarrow$Mn4
distance becomes larger than the Mn4$\leftrightarrow$Mn5 distance.
This is a pure strain gradient related effect, this non-uniformity
is not present in the ground state and in the case of the uniform
strain. This results in the increase of the exchange integral,
$J(\textbf{r}_{4,5})$ in the Heisenberg Hamiltonian between Mn4 and
Mn5 atoms (see Fig.~\ref{4-domains}, bottom panel) due to the
increasing overlap of their wave functions and at the same time in
the decrease of the $J(\textbf{r}_{1,4})$ between Mn1 and Mn4 atoms.
This mechanism is responsible for the out-of-plane rotation of the
Mn4 (and other "inside of the domain" Mn atoms, such as Mn3 and Mn7
on the Fig.~\ref{4-domains}, bottom panel) LMM.

We estimate the flexomagnetic coefficient at the strain gradient
value of $\sim$ 0.4\% from Eq.(\ref{flex-mom}) and the magnetization
vs. strain gradient data shown on Fig.~\ref{magn-vs-flexure} as
follows.

$M_{flex} = \nu\frac{\partial\sigma}{\partial x}$;

$\frac{\partial\sigma}{\partial x}$=$\frac{\partial}{\partial
x}[\frac{\Delta a_{0}}{a_{0}}]$=$\frac{\partial}{\partial
x}[\frac{a_{0}+c\cdot x}{a_{0}}]$=$\frac{1}{a_{0}}c\approx
\frac{0.004}{3.9\cdot 10^{-10}m}$;

$\nu=\frac{M_{flex}}{\frac{\partial\sigma}{\partial
x}}$=$\frac{0.002 \mu_{B}}{0.004/3.9\cdot 10^{-10}m}$ $\Rightarrow
\nu \sim 1.95 \mu_{B}$\AA.

Here $a_{0}=3.9\cdot 10^{-10}m$ is the ground state lattice constant
of the Mn$_{3}$GaN, $M_{flex}=0.002 \mu_{B}$ is the induced
out-of-plane magnetization value at $\sim$ 0.4\% of the strain
gradient. Since the magnetization is linear, the $\nu$ will have the
same value over the considered range of the strain gradient.

In summary, in this paper we discussed the appearance of the net
magnetization in Mn-based antiperovskite compounds (in particular,
Mn$_{3}$GaN) under external strain gradient (\textit{flexomagnetic
effect}). The magnetization dependance on the \textit{flexure} is
linear. The estimated flexomagnetic coefficient is 1.95 $\mu_{B}$\AA
\hspace{0.05cm}. Besides being of substantial theoretical value, the
flexomagnetic effect can have interesting practical applications,
such as in electrical control of magnetization in memory cells if
the structural distortion in the flexomagnetic phase is induced and
can be controlled by the external electric field (i.e. by forming
heterostructure with ferroelectric/piezoelectric compounds).

This work was supported by the National Science Foundation and the
Nanoelectronics Research Initiative through the Materials Research
Science and Engineering Center at the University of Nebraska. This
work was completed utilizing the Blackforest Cluster Computing
Facility of the College of Information Science and Technology at
University of Nebraska at Omaha.

\end{document}